\documentclass[reprint,amsmath,amssymb,aps,prl,amssymb,superscriptaddress]{revtex4-2}
\usepackage[utf8]{inputenc}
\usepackage{bm}
\usepackage{color}
\usepackage{amsmath, braket}
\usepackage{mathtools}
\usepackage{bm}
\usepackage{amsfonts}
\usepackage{dcolumn}
\usepackage{natbib}
\usepackage{bbold}
\usepackage{soul}
\usepackage[dvipsnames]{xcolor}
\usepackage{amssymb}
\usepackage[breaklinks=true,colorlinks,citecolor=blue,linkcolor=blue,urlcolor=blue]{hyperref}
\usepackage{graphicx}
\usepackage{appendix}
\usepackage{soul}

\usepackage{mathtools}
\begin{document}
\title{{Highly}  Efficient Non-relativistic Edelstein effect in $p$-wave magnets}

% Possible author list: Atasi Chakraborty, Rodrigo Jaeshchke Ubiergo, Anna Birk Hellens, Thomas Jungwirth, Jairo Sinova, Libor Smejkal 
\author{Atasi Chakraborty} 
\email{atasi.chakraborty@uni-mainz.de}
\affiliation{Institut f\"{u}r Physik, Johannes Gutenberg Universit\"{a}t Mainz, D-55099 Mainz, Germany}

\author{Anna Birk Hellenes}
\affiliation{Institut f\"{u}r Physik, Johannes Gutenberg Universit\"{a}t Mainz, D-55099 Mainz, Germany}

\author{Rodrigo Jaeschke-Ubiergo}
\affiliation{Institut f\"{u}r Physik, Johannes Gutenberg Universit\"{a}t Mainz, D-55099 Mainz, Germany}

\author{Tom\'{a}s Jungwirth}
%\email{tomas.jungwirth@nottingham.ac.uk}

\affiliation{Institute of Physics, Academy of Sciences of the Czech Republic, Cukrovarnick\'{a} 10, 162 00 Praha 6, Czech Republic}
\affiliation{School of Physics and Astronomy, University of Nottingham, NG7 2RD, Nottingham, United Kingdom}

\author{Libor \v{S}mejkal}
%\email{lsmejkal@uni-mainz.de}
\affiliation{Max Plank Institute for the Physics of Complex Systems, N\"{o}thnitzer Str. 38, 01187 Dresden, Germany}
\affiliation{Institut f\"{u}r Physik, Johannes Gutenberg Universit\"{a}t Mainz, D-55099 Mainz, Germany}
\affiliation{Institute of Physics, Academy of Sciences of the Czech Republic, Cukrovarnick\'{a} 10, 162 00 Praha 6, Czech Republic}

\author{Jairo Sinova}
\email{sinova@uni-mainz.de}
\affiliation{Institut f\"{u}r Physik, Johannes Gutenberg Universit\"{a}t Mainz, D-55099 Mainz, Germany}
\affiliation {Department of Physics, Texas A \& M University, College Station, Texas 77843-4242, USA}

\begin{abstract}
 % ---- Abstract: \textit{[150 words]} -----

The origin and efficiency of charge-to-spin conversion, 
known as the Edelstein effect (EE), has been typically linked to spin-orbit coupling mechanisms, which require materials with heavy elements within a non-centrosymmetric environment. 
Here we demonstrate that the {high}  efficiency of spin-charge conversion can be achieved even without spin-orbit coupling in the recently identified %odd-parity
%non-collinear 
coplanar $p$-wave magnets.
The non-relativistic Edelstein effect (NREE) in these magnets exhibits a distinct phenomenology compared to the relativistic EE, characterized by a strongly anisotropic response and an out-of-plane polarized spin density resulting from the spin symmetries.
%spin symmetry protected out-of-plane polarization 
%since the generated non-equilibrium spin-density polarization is out of the plane, as mandated by the spin symmetries of these $p$-wave magnets.
%reaching one order of magnitude larger than previously reported values in relativistic and other magnetic systems.
% , reaching one order of magnitude larger than previously reported values, can be achieved without spin-orbit coupling. 
% We show that the recently identified odd-parity non-collinear $p$-wave magnets have record values of spin-charge conversion efficiency 
% %can produce significant spin-accumulation 
% even without spin-orbit coupling. 
We illustrate the NREE through minimal %model and $\mathcal{T}t$ symmetric 
tight-binding models, allowing a direct comparison to different systems. 
% we show that NREE in systems with %antisymmetric 
% $p$-wave spin polarization texture surpasses spin susceptibility found in conventional relativistic 2DEG Rashba systems or non-relativistic non-coplanar 3Q (??? antiferromagnets. 
%By spin group symmetry analysis, 
Through first-principles calculations, we further identify  the $p$-wave candidate material CeNiAsO as a %potential
high-efficiency NREE material, 
revealing a 25 times larger response than the maximally achieved relativistic EE and other reported NREE in non-collinear magnetic systems with broken time-reversal symmetry.
%the recently reported NREE in non-collinear %non-coplanar magnet LuFeO$_3$, 
This highlights the potential for efficient spin-charge conversion in $p$-wave magnetic systems.
\end{abstract}

\maketitle

\begin{figure}[t!]
\begin{center}
 \includegraphics[width=\columnwidth]{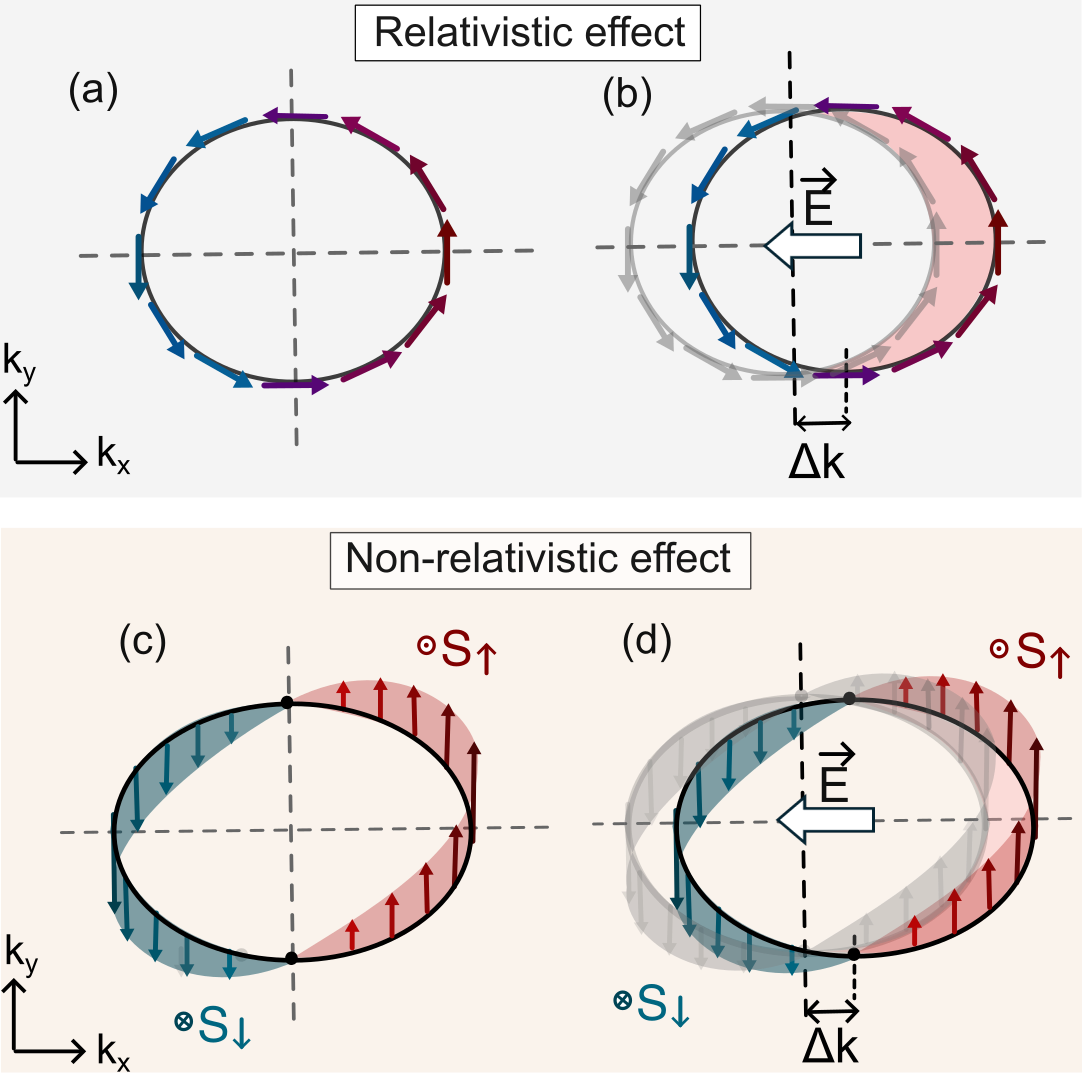}
 \caption{(a) Schematic of a relativistic Rashba spin-texture (1 band). (b) (Relativistic Edelstein effect) Non-equilibrium redistribution resulting in a net in-plane spin-accumulation. (c) Schematic of non-relativistic out-of-plane spin polarization texture of $p$-wave magnets originating from non-collinear $\mathcal{T}\vec{t}$ co-planar magnetic order. (d) (Non-relativistic Edelstein effect) Non-equilibrium out-of-plane spin-accumulation driven by an external electric field.}
  \label{fig:figure1}
 \vspace{-1 cm}
 \end{center}
\end{figure}

%\section{Introduction }
%\begin{center}
%    \textit{[600 words]}
%\end{center}
%\noindent- \textit{Define Edelstein effect and the usual notion of relation with spin-orbit coupling:} 
\noindent Conventional spintronics~\cite{Zutic2004,Maekawa2007} typically relies on ferromagnetic materials and external magnetic fields to generate spin-polarized charge currents. More recent spintronic device concepts focus on  effects generating spin-current or spin-density accumulation by electric fields \cite{Sinova2015,Manchon2019}. The latter is known as the Edelstein effect (EE), %shown in Fig.~\ref{fig:figure1} a and b, 
%or inverse spin-galvanic effect, 
appearing in non-centrosymmetric materials and usually originating from relativistic (spin-orbit coupling) effects, as illustrated in Fig.~\ref{fig:figure1}a and~\ref{fig:figure1}b \cite{Aronov1989,Edelstein1990, Gambardella2011}. 
%
%
%exploit the spin-charge conversion in non-centrosymmetric materials without intrinsic magnetism through an external electric field,
%However, a promising breakthrough called the Edelstein effect (EE) or inverse spin-galvanic effect, offers a compelling alternative by generating spin current in non-centrosymmetric materials without intrinsic magnetism through an external electric field, leveraging spin-orbit coupling (SOC) induced spin degeneracy lifting~\cite{Aronov1989,Edelstein1990, Gambardella2011}. 
%
%
%
%
%
Archetypal applications of the EE include spin-orbit torque  devices, such as the  magneto-electric spin-orbit (MESO) transistor \cite{Manchon2019,Lesne2016,Vaz2019,Manipatruni2019}, where efficient  charge-to-spin conversion is required to exert a torque that can switch the magnetization orientation~\cite{Zelezny2014,Bernevig2005,Manchon2008,Manchon2009,Matos-Abiague2009c,Chernyshov2009,Garate2009}. 
%Spin-charge inter-conversion through relativistic spin-splitting 
The relativistic EE has been demonstrated in numerous systems including Rashba-Dresselhaus systems~\cite{Bryksin2006,Leiva-Montecinos2023}, topological insulators~\cite{Rojas-Sanchez2016a}, semiconductors~\cite{Kato2004,Peters2018}, Weyl semimetals~\cite{Johansson2018,Yang2021a}, oxide interfaces~\cite{Rojas-Sanchez2013b,Shao2016,Trama2024}, 2D electron gases~\cite{Lazrak2023,Johansson2021} and non-centrosymmetric superconductors~\cite{Chirolli2022}. 
%The EE also plays a key role in the proposed concept of 
%Moreover, this electric generation and control of spin in solid-state systems recently became the central theme of next-generation spintronics after the latest discovery of 
%magneto-electric spin-orbit (MESO) transistor, 
%requiring an efficient spin-charge interconversion~\cite{Lesne2016,Vaz2019,Manipatruni2019}.
The non-equilibrium spin density in these systems originates from spin-orbit coupling (SOC), necessitating heavy elements to reach functional efficiencies of charge-to-spin conversion~\cite{Shao2016}.

%corroborated with crystal field-induced spin splitting within non-centrosymmetric systems.
%As a relativistic effect, efficient charge-to-spin conversion necesitates restricted primarily to heavy metal systems~\cite{Shao2016}. %This prompts the natural question: can we identify EE without SOC? 
The possibility of a non-relativistic EE (NREE) was proposed recently in 
%Until very recently, it was proposed that 
non-collinear %non-coplanar 
magnetic systems with broken time-reversal symmetry (TRS)   band structure and broken parity~\cite{Hayami2020b, Naka2019,Gonzalez-Hernandez2023a, Hu2024}, allowing for a %n observable 
current-induced spin accumulation in systems also with possibly light elements. However, these magnetic  %previously considered 
systems do not fully mimic the relativistic EE, since they break TRS %($\mathcal{T}$)  
in momentum space.   
%enabling the generation of non-equilibrium spin current through an applied electric field. 
The recently predicted %co-planar 
p-wave magnets (Fig.~\ref{fig:figure1}c), which preserve TRS in momentum space \cite{Hellenes2023}, provide a direct non-relativistic analog to the relativistic EE.

The prediction and the identification of the favorable characteristics for high NREE efficiency of p-wave magnetic candidates were made possible by applying the spin symmetries approach,  instrumental in the discovery of altermagnets~\cite{Smejkal2021a}. Using the spin symmetries to fully classify and delimit all collinear spin arrangements on crystals leads to the conclusion that only even-parity non-relativistic spin-split band structures (s-, d-, g-, or i-wave) are possible in collinear magnetic systems ~\cite{Smejkal2021a,Smejkal2022a}. %, as we find in Fig.~\ref{fig:figure1}a and c.
This 
% recent discovery of altermagnets, 
% %The recent discovery of altermagnets, 
% based on applying spin symmetries to fully classify and delimit all collinear spin arrangements on crystals. This demonstrated that only even-parity non-relativistic magnetism (s, d, g, or i-wave) is possible in collinear systems ~\cite{Smejkal2020,Smejkal2021a,Smejkal2022a},
%as we find in Fig.~\ref{fig:figure1}a and c, hence 
narrows
the search for $p$-wave (and higher odd-parity) magnets to non-collinear systems \cite{Hellenes2023}.  
%yielded a full classification and delimitation of all collinear spin arrangements on crystals based on spin symmetries. 
%This restrict magnetic order beyond in collinear systems only arise in even-parity (d, g, or i -wave parity) ~\cite{Smejkal2020, Smejkal2021a,Smejkal2022a}. 
%Since p-wave (and higher odd-parity) magnetism is excluded from collinear systems, the search for materials candidates must turn to non-collinear systems  \cite{Hellenes2023}. 
A $p$-wave magnet must break %requires having 
both parity ($P$) and  $P\mathcal{T}$ symmetry, and  %to allow for p-wave spin-texture in momentum space. 
preserve $\mathcal{T}\vec{t}$ symmetry,  where $\vec{t}$ is a translation. This last symmetry preserves TRS in momentum space and enforces zero net magnetization. {P-wave magnetic candidates with collinear polarized band structure have been identified in the co-planar magnets, which have the spin symmetry $[C_{2\perp}||\vec{t}\,\,]$, 
where $C_{2\perp}$ being a 180$^\circ$ spin rotation along the axis perpendicular to the spins  \cite{Hellenes2023}. }
%spin symmetry of [$C_2 || \vec{t}$], combining a 180$^\circ$ spin rotation along the axis perpendicular to the spins and a translation \cite{Hellenes2023}. 
%Very importantly, the combination of these two last 
This spin symmetry mandates that the spin polarization axis in the electronic structure is perpendicular to the plane of the spins. %\st{in crystal space.} 
This sets the NREE (as shown in Fig.~\ref{fig:figure1}d) apart from the usual Rashba  EE, whose non-equilibrium spin polarization is on the plane. It also sets the expectation for a large  NREE efficiency due to the lack of directional averaging of the spin states contributing to the effect, and the much larger spin-splitting (almost 2 orders of magnitude larger) relative to the relativistic EE ~\cite{Ishizaka2011, Bahramy2011}. 

%Contrary to %the even-parity collinear ferromagnets~\cite{Kumar2023, Ahamed2024} or 
%the recently discovered altermagnets~\cite{Smejkal2020, Smejkal2021a, Smejkal2022a, Chakraborty2024, Reimers2024}, which exhibit only even-parity (d, g, or i) magnetism, 
%anti-symmetric collinear $p$-wave spin-polarization texture can only be observed in non-collinear magnets due to the presence of $C_{2\perp}\mathcal{T}$ in combination with $\mathcal{T}t$ symmetry.  

%Since the generation of the spin accumulation is dictated by the exchange splitting of the system, the non-relativistic spin splitting of $p-$wave magnets can surpass by $2$~orders than the largest SOC-driven  Rashba splitting of $\sim10^2$~meV observed in BiTeI~\cite{Ishizaka2011,Bahramy2011}. This raises the expectations for a larger spin-to-charge conversion efficiency.   
%
%Analysis of the Kubo linear response expression for non-equilibrium spin-density suggests the intra-band susceptibility containing the Fermi surface contribution only survives in $\mathcal{T}\vec{t}$ symmetric $p$-wave magnets.  
%
%To establish the feasibility of electric-field induced non-equilibrium spin accumulation, 
We first demonstrate this physics in 
%We begin with exploring 
a simple 2D generic minimal model that exhibits a $p$-wave spin-polarization band structure.
%We formulate a simple 2D minimal model to check the feasibility of electric field-induced non-equilibrium spin-accumulation in the $p$-wave spin-polarization texture. 
We find, even at the simple model level, that the NREE susceptibility of $p$-wave magnets surpasses the magnitude calculated for the relativistic Rashba two-dimensional electron gas (2DEG) and recently reported non-relativistic non-coplanar $3Q$ anti-ferromagnets (AFMs) with broken TRS~\cite{Edelstein1990, Johansson2016, Gonzalez-Hernandez2023a}. 
%A comparison of the Rashba spin texture, generating relativistic EE, and the $p$-wave spin polarization texture, promoting NREE, is shown in Fig.~\ref{fig:figure1}d and e.  
Extending our model analysis to a bi-Kagome lattice geometry, we find a substantial increase in spin-accumulation density. Finally, through first principle calculations, we identify the $p$-wave candidate CeNiAsO showing a highly efficient NREE, exhibiting 25 times larger response than the highest Rashba EE \cite{Rojas-Sanchez2016a} and the non-collinear AFM LuFeO$_3$ \cite{Gonzalez-Hernandez2023a}.

\section{Results}
\subsection{Minimal $p$-wave model}
%\begin{center}
%    \textit{[500 words]}
%\end{center}

\begin{figure}[t!]\label{fourband_minimal}
\begin{center}
 \includegraphics[width=\columnwidth]{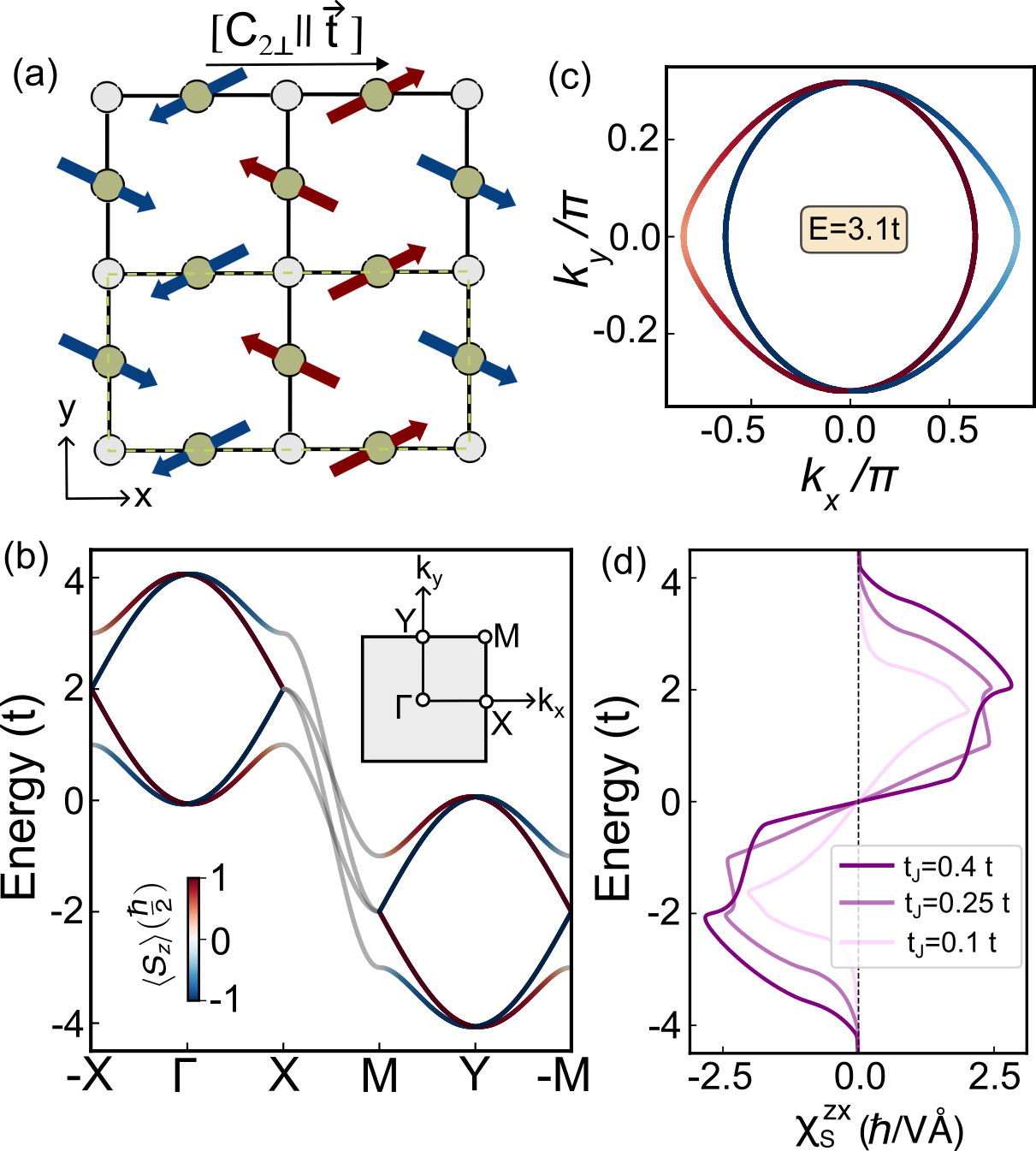}  
 \caption{(a) Schematic of the lattice model with a coplanar non-collinear spin arrangement on the crystal with the spin arrangement that realizes the unconventional p-wave phase. The unit cell is marked with dashed green line. (b) Energy dispersion for $t_J=0.25~t$ along high-symmetry momentum paths  showing parity broken time-reversal symmetric spin splitting. The color represent the out-of-plane spin component. The opposite spin bands are completely degenerate along the $k_y$ direction forming a nodal line. (c) Constant energy $E=3.1t$ contour with $t_J=0.25~t$.  (d) Variation of the NREE response with the strength of the exchange term. An increase of $t_J$ enhances the shifting of opposite spin bands in opposite momentum directions, therefore, the response strength $\chi_S$ increases.}\label{fig:figure2}
 \end{center}
\vspace{-0.5 cm}
\end{figure}

%\noindent \textbf{\textit{Minimal $p$-wave model}:} 
To exemplify the $p$-wave electronic structure and its NREE response, we first formulate {a}  low-energy minimal model for the odd-parity magnets based on the simple tight-binding model presented in Fig.~\ref{fig:figure2}a. Here the opposite spins are connected by $[C_{2\perp}||\vec{t}\,\,]$ symmetry. The {coplanar non-collinear} magnetic ordering doubles the unit cell of the square lattice (shown as a dashed green line in Fig.~\ref{fig:figure2}a). The electron hopping among the grey nonmagnetic sites is parameterized by $t$, and the p-wave spin splitting originates from the exchange-dependent hopping parametrized by $t_J$.
The minimal four-band Hamiltonian is given by ~\cite{Hellenes2023},
\begin{equation}\label{eq:lattice_model}
    H= 2t(\mathrm{cos}\frac{k_x}{2}\tau_1+\mathrm{cos}k_y)+2t_J(\mathrm{sin}\frac{k_x}{2}\sigma_1\tau_2 +\mathrm{cos}k_y\sigma_2\tau_3)~.
\end{equation}
Pauli matrices $\sigma$ and $\tau$ correspond to the spin and site degrees of freedom. The detailed construction of the Hamiltonian is included in SI. Here {$t$, $t_J$ } represent the spin-independent and exchange-dependent hopping terms between the nearest neighbor sites within the square lattice geometry.

We illustrate the band structure calculated at $t_J=0.25t$ parameter choice of Eq.~\ref{eq:lattice_model} in Fig.~\ref{fig:figure2}b~. 
The fixed energy contour plot at energy $E=3.1t$  is plotted in Fig.~\ref{fig:figure2}c. As it is clear from the figures, in this low-energy model, the relative displacement of Fermi surfaces of opposite spin,
results in a spin-split odd-parity-wave band structure. 
%Here the exchange coupling $t_J$ effectively lifts the degeneracy of Kramer's pairs by including a momentum-dependent opposite shift. 
The band-pairs with opposite out-of-plane spin polarization form a nodal crossing along the $k_y$ direction. While the direction of the polarization of the states is out of the plane and protected by the spin symmetries of the model,
we emphasize that the spin-polarization magnitude of $p$-wave magnet is not protected, varying across the Brillouin zone. This $k$-dependent spin polarization gradient is essential for the finite NREE in the single-particle models of spin-split bands related by time reversal (see SI for more details). 

To compute the  non-equilibrium spin density accumulation $\delta \mathbf{S}$ due to an electric field, we use the 
Kubo linear response theory as $\delta S^i=\chi_{S}^{ij}E^j~$, where $\mathbf{E}$ is the %time-independent 
applied electric field. The %dissipative 
spin-current response function has the following expression
\begin{equation}
    \chi_{S}^{ij}= \frac{1}{2\pi} Re \sum_{\mathbf{k}\alpha \beta}S^i_{\alpha \beta}(\mathbf{k}) v^j_{\beta \alpha}(\mathbf{k}) [ G^R_{\mathbf{k}a}G^A_{\mathbf{k}b} - G^R_{\mathbf{k}a}G^R_{\mathbf{k}b} ] ~.
\end{equation}
Here, $G^{R(A)}_{\mathbf{k}a}=1/(\epsilon_F-\epsilon_{\mathbf{k}a} \pm \frac{i\hbar}{2\tau_{\mathbf{k}a}})$ is the retarded (advanced) Green's function evaluated w.r.t the Fermi energy $\epsilon_F$. $\tau_{\mathbf{k}a}$ is the quasiparticle lifetime, taken here to be constant, $\hbar/\Gamma$. We can separate the spin density into two parts depending on the contributions from the intra-band, $\delta S_{\mathrm{intra}}$ and inter-band, $\delta S_{\mathrm{inter}}$. The intra-band term with fermi surface contribution has the following expression~\cite{Garate09,Zelezny2014a}, 
\begin{equation}\label{intra-term}
    \delta \mathbf{s}_{\mathrm{intra}}= \frac{e\hbar}{2\Gamma}\int \frac{d^2k}{(2\pi)^2} \sum_{\alpha} \mathbf{S}_{\mathbf{k}\alpha} (\mathbf{E}\cdot \mathbf{v})_{\mathbf{k}\alpha} \delta(E_{\mathbf{k}\alpha}-E_F)~.
\end{equation}
The $\delta \mathbf{s}_{\mathrm{intra}}$ has the dominant contribution to charge spin conversion. The inter-band term can be expressed as, 

\begin{align}\label{inter-term}
    \delta \mathbf{s}_{\mathrm{inter}}= e\hbar\int \frac{d^2k}{(2\pi)^2} &\sum_{\alpha\ne \beta} (f_{\mathbf{k}\alpha}-f_{\mathbf{k}\beta}) \mathrm{Im}[\mathbf{S}_{\alpha\beta} (\mathbf{E}\cdot \mathbf{v})_{\beta\alpha}]\nonumber \\
     &\times  \frac{(E_{\mathbf{k}\alpha}-E_{\mathbf{k}\beta})^2-\Gamma^2}{[(E_{\mathbf{k}\alpha}-E_{\mathbf{k}\beta})^2-\Gamma^2]^2}~.~
\end{align}
%

%%%%%%%%%%%% New Figure 2 %%%%%%%%%%%%%%%%
%%%%%%%%%%%%%%%%%%%%%%%%%%%%%%%%%%%%%%%%% 

Under time reversal symmetry (TRS) $S \rightarrow - S$ and $\mathbf{v} \rightarrow -\mathbf{v}$. 
%and $\tau \rightarrow -\tau$. 
Therefore,  $\delta S_{\mathrm{intra}}$ is allowed under TRS. However, the kernel of $\delta S_{\mathrm{inter}}$ has very similar symmetry properties as the Berry curvature of the system, which vanishes if the system preserves TRS. For this reason, the `intra' and `inter' band components are often referred as $\mathcal{T}^{even}$ and $\mathcal{T}^{odd}$ part of the Edelstein response tensor. As $\mathcal{T}^{odd}$ is the odd Fermi surface property under time-reversal symmetry, it is only allowed for ferromagnets, altermagnets and some non-collinear AFMs breaking TRS %such as ANMn$_3$
~\cite{Gurung2021a, Gonzalez-Hernandez2023a} but prohibited in the $p$-wave candidates,  as we also verify in our numerical calculations. 
For the identified $p$-wave material candidates in Ref.~\cite{Hellenes2023} from the materials listed in {\small MAGNDATA}~\cite{Gallego2016}, we have included the symmetry-allowed tensorial form of the susceptibility in Table-I of SI.

%Here by combining the minimal lattice model and Kubo linear response theory, we calculate the Edelstein susceptibility, $\chi_S^{ij}\equiv~\delta S^i/ E^j$, for different values of exchange-coupling $t_j$. Here $\delta S$ and $\mathbf{E}$ represent spin density and applied electric field, respectively. 

For our minimal model Hamiltonian, the $\chi^{zx}_S$ is the only finite component of the {symmetry-constrained rank-2 tensor}  (Fig.~\ref{fig:figure2}d).  The maximum value of the $\chi_S$ increases with $t_J$ due to the increasing anisotropy between the spin channels. 
The particle-hole symmetry evidenced in the response (Fig.~\ref{fig:figure2}d) is in agreement with the model band structure shown in Fig.~\ref{fig:figure2}b.
%\textcolor{red}{The symmetric nature of the NREE for positive and negative energy reflects the inherent particle-hole symmetry of the band dispersion}.
In order to compare to other systems, we look at  $\chi_S$ within 
%For the sake of comparison of $\chi_S$ strength within a model level, we choose two examples: 
(i) a non-coplanar AFM model with broken TRS with $3Q$ spin-texture ~\cite{Gonzalez-Hernandez2023a} and (ii) a two-dimensional electron gas (2DEG) with relativistic Rashba SOC.  For the former case, the non-equilibrium susceptibility, integrated over the unit cell, is reported to be $\chi_S^{3Q-AFM}\sim0.5 \hbar \mathrm{\AA}/V $~\cite{Gonzalez-Hernandez2023a}. The non-equilibrium spin density for the Rashba 2DEG is given by  $\delta S^R/eE =~\alpha_R m_0/\hbar^2 \pi \Gamma$~\cite{Edelstein1990}.
We set $\Gamma=0.1$~eV, $\alpha_R=10^{-9}$ eVm as the typical Rashba strength expected in transition metal heterostructures ~\cite{Manchon2015, Johansson2016}, with $m_0$ being the free electron mass.
This gives an effective Edelstein response  $\chi_S^{R}\sim0.04 \hbar /V \mathrm{\AA}$. 
Already at the simple model level, the strength of $\chi_S$ of the minimal $p$-wave model is 1-2 orders larger than that of the reported non-relativistic non-coplanar 3Q-AFM and relativistic Rashba 2DEG model.

%\begin{equation}
%    H=\mathbf{\sigma}_o\otimes \sum_{\mathbf{k},\sigma} E_{\sigma}(\mathbf{k}) c^\dagger_{\mathbf{k}\sigma} c_{\mathbf{k}\sigma}~+\mathbf{\sigma}_z\otimes J_s\sum_{\mathbf{k},\sigma\ne \sigma^\prime}c^\dagger_{\mathbf{k}\sigma} c_{\mathbf{k}\sigma^\prime},
%\end{equation}
%\begin{equation}\label{formEk}
% \mathrm{with},~~   E_{\sigma}(\mathbf{k})= \sum_{\alpha} (-2t\mathrm{cos}k_{\alpha}-2\sigma Jb_{\alpha}\mathrm{sin}k_{\alpha})~.
%\end{equation} 
%Here, $c^\dagger_{\mathbf{k}\sigma}$ and $c_{\mathbf{k}\sigma}$ are the fermionic creation and annihilation operators for momentum $\mathbf{k}$ and spin $\sigma$. The band dispersion for the minimal model with $J_s=0$ along the high-symmetry paths is plotted in Fig.~\ref{fourband_minimal}. We choose $b_x=0.2 t/J$ and $b_y=0.0$ for the plot. Here with $J_s=0$, for the non-zero value of any $b_{\alpha}$, the spin bands displaced with opposite offset at the origin preserving $E_{\sigma} (\mathbf{k})=~E_{-\sigma}(-\mathbf{k})$.  While from eqn.~\ref{formEk}, we obtain $E_{\sigma}(\mathbf{k})\ne~E_{\sigma}(-\mathbf{k})$. Consequently, the Hamiltonian satisfies the criteria of broken parity and time-reversal invariance similar to $^3$He.

%%%%%%%%%% Figure 3 %%%%%%%%%%%%%%%%%%%%
\begin{figure*}
\begin{center}
 \includegraphics[width=1.8\columnwidth]{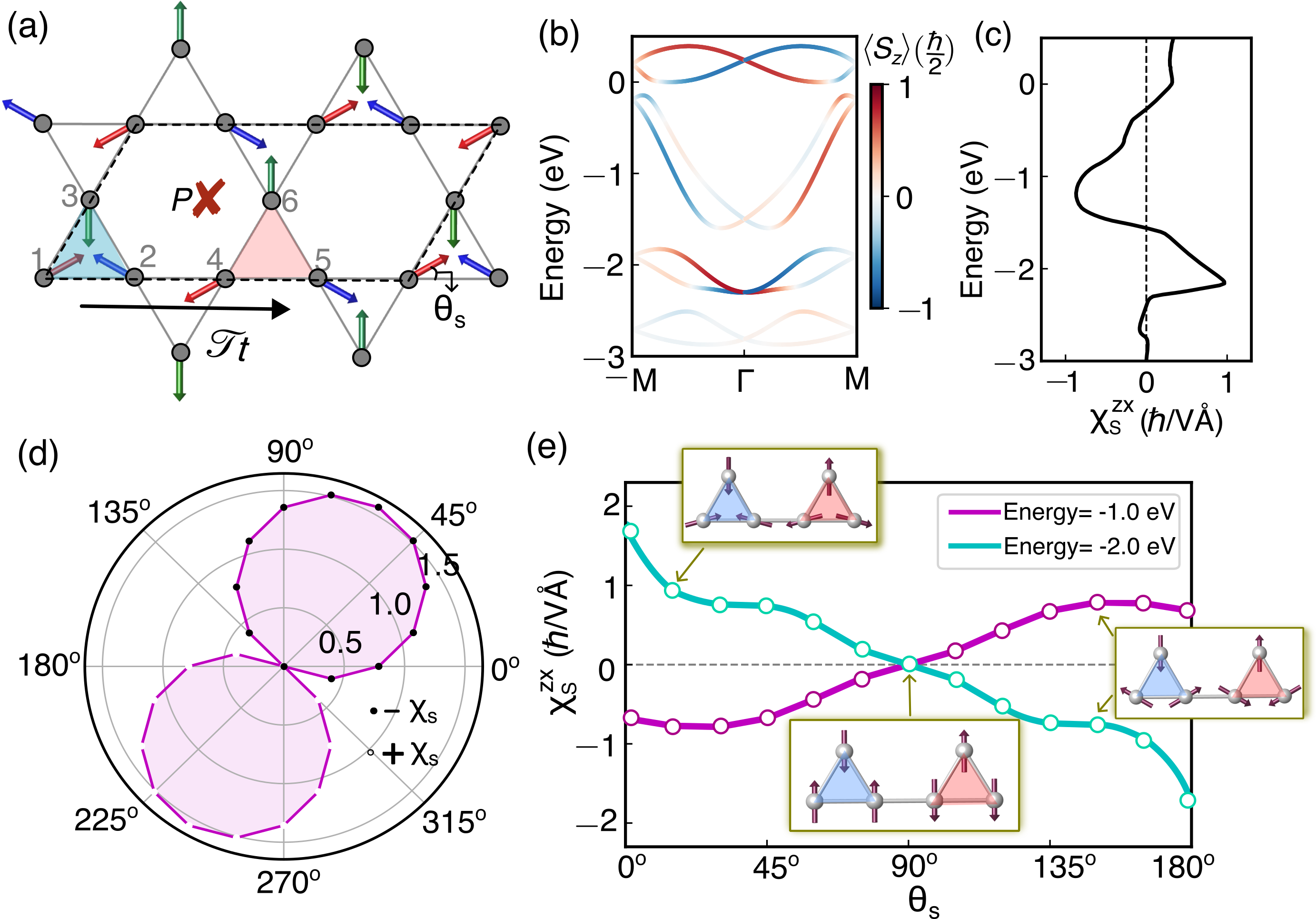}
 \caption{(a) Direct space magnetic order of the Kagome lattice with twice propagation along the $x$-axis. The magnetic order breaks the inversion and exhibits combined TRS and lattice translation symmetry, $\mathcal{T}\vec{t}$ marked by the black arrow, which connects the spins in the blue and red shaded triangles. The coplanar spins in each magnetic triangle observe $120^\circ$ spin alignment. The spin canting angle of the red spin, $\theta_s$, is also marked. %Similarly, the canting of the blue spin within the same triangle can be defined as $(180^\circ-\theta_s)$. 
 (b) The out-of-plane spin-projected band dispersion along $-M \mbox{-} \Gamma \mbox{-} M$ high-symmetry direction. (c) The non-equilibrium  intra-band susceptibility density, $\chi_S^{zx}$, for electric field $\mathbf{E}= E\hat{x}$. (d) The angular distribution of $\chi_S^{zx}$ (in units of $\hbar /V\text{\AA}$) for energy -0.1~eV w.r.t  the electric field direction. The white and black circles represent the positive and negative sign of $\chi_S$ (e). The magnitude of $\chi_S^{zx}$ gradually increases with the spin canting angle $\theta_s$, relative to the collinear arrangement at $\theta_s=90^\circ$.
 %For the collinear configuration, i.e. $\theta_s=90^\circ$ the bands are not spin split; hence, the non-equilibrium spin density vanishes as per expectation.
 }\label{fig:figure3}
 \end{center}
\end{figure*}
%%%%%%%%%%%%%%%%%%%%%%%%%%%%%%%%%%%%%%%%% 

%\vspace{2 cm}
\subsection{Tight Binding model of $p$-wave bi-kagome {magnet}}
%\begin{center}
%    \textit{[500 words]}
%\end{center}

%\noindent- \textit{Model Hamiltonian and symmetry analysis:} 
%##############################################################
We next analyze the NREE in $p$-wave magnets within a multi-orbital two-dimensional Kagome lattice, as described by
%by employing a multi-orbital tight-binding (TB) model Hamiltonian with magnetic exchange within a two-dimensional Kagome lattice, as described by 
the Hamiltonian,
\begin{equation}\label{eqn_bikagome}
    H= \sum_{i} \bm{\mathcal{J}}_i\cdot \bm{\sigma}~c_i^\dagger c_i + t_h\sum_{<ij>}c_i^\dagger c_j~.
\end{equation}
Here, $c^\dagger$ and $c$ are the fermionic creation and annihilation operators. $i,j$ are the site indices. The first {term in eq.~\ref{eqn_bikagome}} represents the interaction term {where} $\bm{\mathcal{J}}_i$ as the local exchange parameter. $\bm{\sigma}$ is the vector containing three spin Pauli matrices. The kinetic energy of the Hamiltonian is included in the second term, where $t_h$ is the isotropic nearest neighbor hopping strength. We choose the spin direction for six-inequivalent sites to be ${\hat{\mathcal{J}}_1}=-{\hat{\mathcal{J}}_4}=(\mathrm{cos}\theta_s ~\hat{x}+\mathrm{sin}\theta_s~\hat{y})$, ${\hat{\mathcal{J}}_2}=-{\hat{\mathcal{J}}_5}=(-\mathrm{cos}\theta_s~\hat{x}+ \mathrm{sin}\theta_s~\hat{y})$, ${\hat{\mathcal{J}}_3}=-{\hat{\mathcal{J}}_6}=-\hat{y}$ with $\theta_s=\frac{\pi}{6}$.  The coplanar noncollinear magnetic ordering shown in Fig.~\ref{fig:figure3}a fulfils the spin symmetry criteria for collinear p-wave spin-polarization band structure presented earlier. The neighboring red and blue triangular sub-units containing opposite $\Gamma_{4g}$ phases~\cite{Gurung2021}, are connected by $\mathcal{T}\vec{t}$ symmetry as shown by the black arrow. Although the $120^\circ$ spin-order preserves the parity for a single Kagome unit cell, the  $\mathcal{T}\vec{t}$ symmetric spin-order on the Kagome lattice breaks the inversion.  As already emphasized before, due to the  $[C_{2\perp}||\vec{t}\,\,]$ spin symmetry, the in-plane $S_x$ and $S_y$ components of the spins are identically zero {for every}  momentum. The $\langle S_z \rangle$ polarized band structure along the high-symmetry $-M \mbox{-} \Gamma \mbox{-} M$ direction is shown in Fig.~\ref{fig:figure3}b. Here we choose $|\bm{\mathcal{J}}|=|t_h|=$1.0 eV throughout our model calculations.

Our calculations reveal a significant NREE for the Kagome TB model. We plot the $\chi_S$ response choosing the direction of the applied electric field along $x$ with $\Gamma=0.01$ eV in Fig.~\ref{fig:figure3}c. The strength of the NREE susceptibility is of the same order as {in} the minimal $p-$wave model. 
We also show in Fig.~\ref{fig:figure3}d the expected $p$-wave anisotropy in the 
%To explore the 
directional dependence of $\chi_S$, by rotating the $\vec{E}$ in the $x-y$ plane and plotting the $\chi_S$ for a fixed energy $E-E_F=-0.1$ eV in Fig.~\ref{fig:figure3}d. The black and white circles represent the positive and negative signs of the NREE susceptibility. The $\chi_S$ shows a nodal line for $\mathbf{E}||(-\frac{\sqrt{3}}{2} \hat{x}+\frac{1}{2}\hat{y}$) independent of the chemical potential. 
Hence, the angular measurement of the NREE provides a strong measurable signature for this NREE in these $p$-wave magnets. 
%essentially captures the information on the distribution of the pole of the out-of-plane spin-polarization over the electronic bands. 

We next explore the NREE dependence on the chirality of spins within the Kagome lattice, by adjusting the canting angle between the spins $\theta$ (inset of Fig.~\ref{fig:figure3}a and e). As a result, the $C_3$ symmetry of neighboring spins within individual triangles is disrupted, while the overall $[C_2||\vec{t}\,\,]$ connecting the red and blue triangles is preserved. The change in spin canting configuration changes the $\chi_S$ anisotropy pattern from that of the $120^\circ$ order (see SI for more details).  
For $\theta_s=90^\circ$, $\chi_s$ vanishes since the spin-order becomes collinear promoting spin degenerate bands, consistent with the result from the full classification and delimitation of collinear spin arrangements on lattices by the spin symmetries
~\cite{Smejkal2021a}. %Therefore, $\chi_s$ vanishes for $\theta_s=90^\circ$. 
In Fig.~\ref{fig:figure3}e we have plotted the NREE for two different energy values for $\theta_s$ in the range of $0^\circ-180^\circ$. The increase in canting angle relative to the collinear arrangement ($\theta_s=90^\circ$) enhances the NREE and changes  sign at $\theta_s=90^\circ$.

\begin{figure*}
\begin{center}
 \includegraphics[width=2\columnwidth]{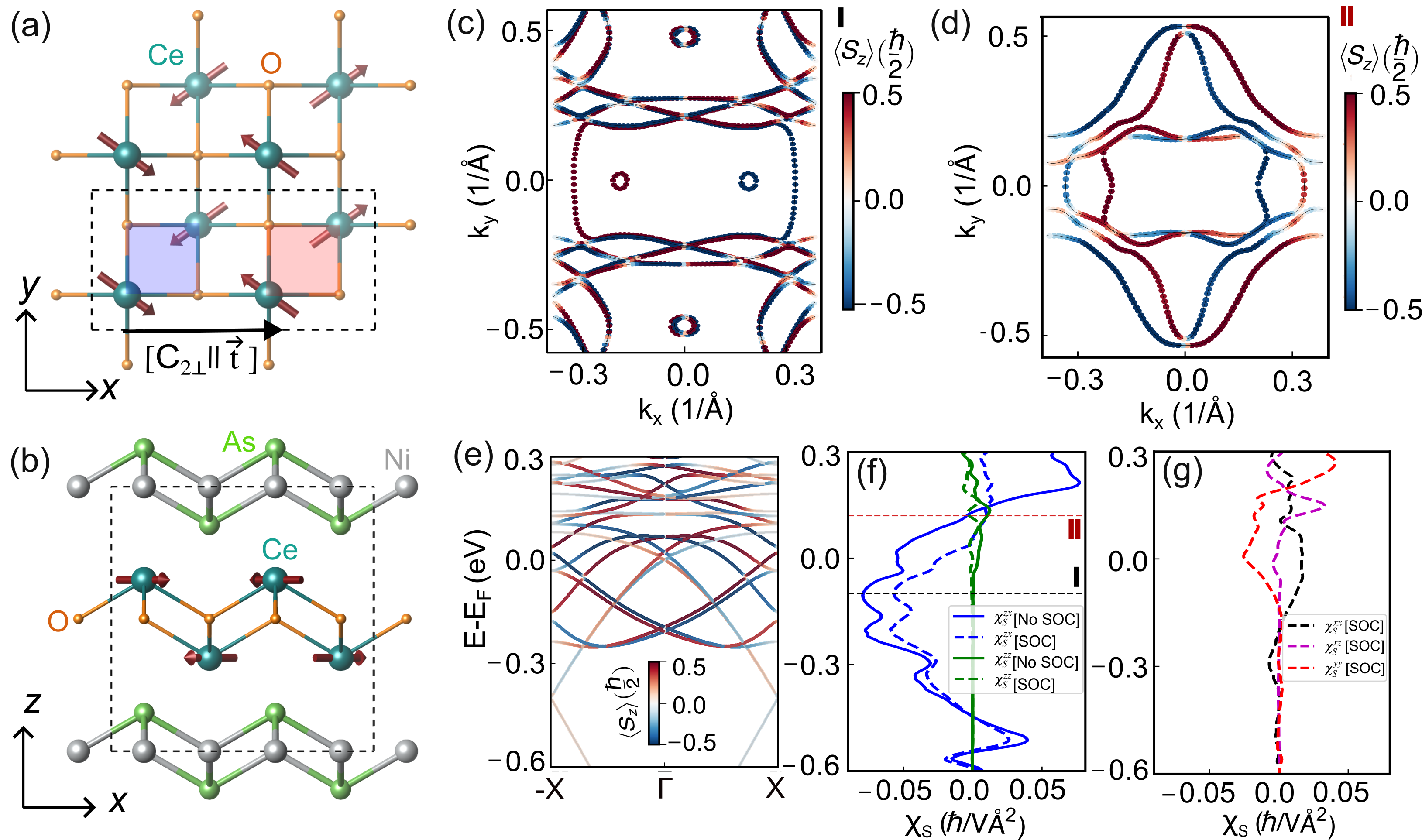}
 \caption{ Top (a) and side (b)  view of the unit cell of CeNiAsO crystal and the co-planar magnetic order with $\mathcal{T}t$ symmetry. 
 The out-of-plane spin projected constant energy isosurface at $E\sim -0.1$ eV (c) and $E\sim 0.12$ eV (d) in the  $k_z=0$ momentum plane. 
 (e) Nonrelativistic odd-parity wave spin splitting of energy bands plotted along the $-X\mbox{-}\Gamma \mbox{-}X$ path. The colormap represents the out-of-plane $\langle S_z \rangle$ component. 
 (f) The symmetry-allowed components of non-equilibrium intra-band susceptibility density, $\chi_S^{zx}$ and $\chi_S^{zz}$.  The modifications of the specific components in the presence of SOC are shown in dashed curves. The isoenergy cuts of (c) and (d) are at the energy where the $\chi_S^{zx}$ component of the NREE susceptibility acquires maximum and minimum values as indicated by dotted lines in (f). (g) Susceptibility components arise due to the sole relativistic effect. SOC leads to finite $\chi_S^{xx}$, $\chi_S^{xz}$ and $\chi_S^{yy}$ which are not allowed within the NREE susceptibility tensor. 
%\textcolor{red}{New version with full band dispersion and vertically plotted $\chi$. This looks very much busy to me.}
%
%\textcolor{red}{[AC: If we need to show the bands in the full energy scale like the model figure, we need to include the whole path similar to the model to support the contributions in $\chi$, which will make the figure more clumsy. Therefore keeping the older energy scale.] } %The angular distribution of $\chi_S$ (in units of $\hbar /V\text{\AA}^2$) for energy $-0.2$~eV w.r.t changes the electric field direction in the $x-y$ plane. The purple and grey circles represent the sign of the $\chi_S$.
 } \label{fig:figure4}
 \end{center}
\end{figure*}
%%%%%%%%%%%%%%%%%%%%%%%%%%%%%%%%%%%%%%%%%

\subsection{Material Candidate: $\mathbf{\mathrm{CeNiAsO}}$}

%\subsection{$p$-wave magnet: CeNiAsO}
%\begin{center}
%    \textit{[500 words]}
%\end{center}
%\noindent- \textit{Spin symmetry of CeNiAsO}\\
%\textit{From pwave arXiv--} 

%So far, we have utilized  model systems to confirm the feasibility of the NREE based on fundamental symmetry principles in  $p$-wave magnets. 
%Now to explore the actual quantitative estimation of the realistic non-equilibrium spin accumulation for the $p$-wave candidate material CeNiAsO.  

We next compute the NREE for the realistic $p$-wave magnetic material candidate CeNiAsO \cite{Hellenes2023}.
The system shows a co-planar commensurate magnetic order with a moment 0.37 $\mu_B$ below N\'{e}el temperature, $T_N=7.6$K~\cite{Wu2019}. From the spin symmetry analysis, %presented in section~\ref{sec:symmetry_analysis}, 
the experimentally reported coplanar non-collinear magnetic order of CeNiAsO, as shown in Fig.~\ref{fig:figure4}a and~\ref{fig:figure4}b, 
implies the {coplanar spin only} group symmetry $\boldsymbol{r}_{s} =\lbrace{E,\bar{C}_{2z}\rbrace}$, where $\bar{C}_{2z}$ is $180^\circ$ rotation around $z$-axis combined with spin-space inversion (time reversal).
%, and the spin-rotation $z$-axis is orthogonal to the spin-space $x-y$ plane of the coplanar spin arrangement. 
The nontrivial spin-space group   $\boldsymbol{G}^{S}$ in CeNiAsO contains the following symmetry elements~\cite{Hellenes2023, Smolyanyuk2024b, Shinohara2024}:
$[E||E]$, $[E||\mathcal{M}_{y}\vec{t}_{\frac{b}{2}}]$, $[C_{2z}||\vec{t}_{\frac{a}{2}}]$, $[C_{2z}||\mathcal{M}_{y} \vec{t}_{(\frac{a}{2}+\frac{b}{2})}]$, $[C_{2x}||C_{2y}\vec{t}_{\frac{b}{2}}]$, $[C_{2y}||C_{2y} \vec{t}_{(\frac{a}{2}+\frac{b}{2})}]$, $[C_{2x}||P]$, $[C_{2y}||P \vec{t}_{\frac{a}{2}}]$, 
where $C$, and $\mathcal{M}$ denote rotation and mirror operations, and $\vec{t}_i$ is the translation along $i$ of the lattice vector. The symmetry operation $[C_{2z}||\vec{t}_{\frac{a}{2}}]$ (see Fig.~\ref{fig:figure4}a) in combination with broken inversion symmetry fulfils the above symmetry conditions for the odd-parity-wave magnetic state. % provided in section~\ref{sec:symmetry_analysis}. 
Additionally, the symmetry $[C_{2y}||C_{2y}]$ enforces 
%the existence of the p-wave state with 
a single spin-unpolarized line $k_{x}=k_{z}=0$ in the $k_{z}=0$ plane, implying $E(k_x, k_y, k_z,\sigma_{z})=E(-k_{x}, k_{y}, -k_{z},-\sigma_{z})$.
%
%\noindent- \textit{Band structure and highlight on out-of-plane spin polarization}\\
%To study the effect of ground state magnetic order on spin-polarization texture, 
We plot the band dispersion of CeNiAsO in Fig.~\ref{fig:figure4}e. The bands have opposite out-of-plane spin polarization $S_z$ for opposite momentum with a linear crossing at the $\Gamma$ point. The fixed energy contour at $E-E_F=-0.1$ eV and at $+0.12$ eV within $k_x-k_y$ plane (Fig.~\ref{fig:figure4}c, ~\ref{fig:figure4}d) shows a nodal line along $k_y$ direction.  

We compute all the components of the NREE  susceptibility for metallic CeNiAsO in the limit of zero SOC, shown by solid lines in Fig.~\ref{fig:figure4}f. 
%As only the intra-band component survives in a system with  $\mathcal{T}\vec{t}$ symmetry, the effect is only expected in metallic systems, as is the case for the metallic CeNiAsO.
%the only contributing term in $\chi_S$ for $\mathcal{T}\vec{t}$ system, is a Fermi surface property, we get a finite response at the Fermi energy due to the metallic nature of CeNiAsO. 
Our spin symmetry analysis (see SI for details) yields that only the  $\chi_S^{zx}$ and $\chi_S^{zz}$ components survive, which is consistent with our numerical results. Given the negligible van-der-Waals interaction between the Ce layers stacked along the $z$-axis within the bulk geometry, we find a small $v_z$ component of the velocity. Therefore the obtained $\chi_S^{zz}$ value is substantially lower than that of $\chi_S^{zx}$. {Fig.~\ref{fig:figure4}c and ~\ref{fig:figure4}d represent the isoenergy contours at energies where the NREE susceptibility $\chi_S^{zx}$ reaches the maximum and minimum values, as shown with dotted lines in Fig.~\ref{fig:figure4}f.}

%The polar distribution of $\chi_S$ over different $\mathbf{E}$ orientations shows that $\chi_S$ vanishes for $\mathbf{E}||\hat{y}$ and is a maximum at $\mathbf{E}||\hat{x}$. 
%The NREE susceptibility increases gradually with a deviation of $\mathbf{E}$ from crystallographic $y$-axis and peaks when $\mathbf{E}||\hat{x}$.   

Of course, along with non-relativistic exchange-driven effects, SOC can also have an impact due to the fact that Ce is a rare-earth element. 
Within the relevant magnetic point group analysis, 
%within the relativistic limit. Interestingly, the $C_{2\perp}t$ symmetry operation, 
by its own construction, the spin group symmetry  $[C_{2\perp}||\vec{t}\,\,]$   is broken by spin-orbit coupling and, therefore, no longer enforces the polarization direction to be out of the plane. The induced %(but much weaker) 
finite in-plane spin components hence emerge solely due to SOC effect. 
In Fig.~\ref{fig:figure4}f, we show the change of the NREE susceptibility non-zero components in the presence of SOC with dashed curves.
The susceptibility components emerging solely from the relativistic origin are plotted in Fig.~\ref{fig:figure4}g. 
%the out-plane-spin polarization enforced by the spin symmetry enforcing out-of-plane spin-polarization texture within the non-relativistic regime, is no longer included in the magnetic point group for CeNiAsO. Therefore, we find that the in-plane spin components also emerge solely due to SOC effect. 
The details of relativistic band dispersion and the corresponding tensorial form of the non-equilibrium spin accumulation susceptibility are included in section 5 of SI. 
While the NREE dominates prominently over a large range of energies, the contributions of the SOC susceptibilities of in-plane polarization can be exploited to calibrate the relevance of this contribution in general. 
Comparing the relative magnitude of the in-plane and out-of-plane EE susceptibilities should be a good experimental test of the relevance of the NREE. 
%\noindent- \textit{Non-equilibrium spin accumulation plots.}\\

Next, we compare the magnitude of NREE susceptibility of $\mathcal{T}\vec{t}$ symmetric $p$-wave magnets with recently reported non-relativistic spin accumulation within {non}-collinear magnets with broken TRS, i.e. non $p$-wave magnets. The  NREE reported for non-centrosymmetric LuFeO$_3$ is 0.5 $\hbar \text{\AA}/V$ calculated within unit cell of volume 360.61 $\text{\AA}^3$~\cite{Gonzalez-Hernandez2023a}. The effective $\chi_S^{zx}$ integrated over the unit cell of volume 267.50 $\text{\AA}^3$  of CeNiAsO gives NREE $\sim 13~\hbar \text{\AA}/V$, which is 25 times higher in magnitude than that of NREE reported for LuFeO$_3$.

\section{Discussion and Outlook}
%\begin{center}
%    \textit{[500 words]}
%\end{center}
%\noindent- Summary of the work\\
We predict here a highly efficient NREE in $p$-wave magnets. This type of coplanar noncollinear magnetic order preserves TRS in momentum space (through the $\mathcal{T}\vec{t}$ symmetry) while exhibiting $p$-parity polarized band structure \cite{Hellenes2023}.  Their spin symmetry $[C_{2\perp} ||\vec{t}\,\, ]$ forces the polarization direction in the band structure to be perpendicular to the plane of spin coplanarity. This unique feature of $p$-wave magnets gives them an {anisotropic} EE signature relative to their relativistic and other magnetic system counterparts and can lead to unprecedented high efficiency in spin-to-charge conversion. % through this effect. 
 {The non-equilibrium spin accumulation within $p$-wave magnets does not originate from the SOC effect but rather from the exchange interaction of the non-collinear magnetic order.}
The inevitable presence of SOC can lead to contributions to the EE within these materials, particularly in the presence of heavy elements, {and} its contribution to the 
spin-charge conversion efficiency can be estimated by the measured in-plane vs. out-of-plane susceptibilities.  
However, this contribution is expected to be significantly less {pronounced} in material candidates with lighter elements. 
%This gives a unique opportunity to confirm this effect exprimentally, by comparing the different susceptibilities and their unique anisotropic signatures. 
We predict here that the charge-to-spin conversion in the metallic material candidate CeNiAsO, showing coplanar noncollinear magnetic order, 
should be dominated by the NREE. 
%show a strong NREE observation of this large charge-to-spin conversion.  
%In addition to the NREE, a direct consequence of this unique type of NREE, is a  
The NREE in $p$-wave magnet suggests the possibility of current-driven controlled {spin} dynamics, which is fundamentally different from the relativistic effects observed in conventional spin-transfer and spin-orbit torque devices.

\section{Methods}

%\subsection{First Principles calculation}
We have used density functional theory (DFT) in the plane wave basis set. We used the Perdew-Burke-Ernzerhof (PBE)~\cite{Perdew1996} implementation of the generalized gradient approximation (GGA)  for the exchange-correlation. This was combined with the projector augmented wave potentials~\cite{Blochl1994, Kresse1999}  as implemented in the Vienna {\it ab initio} simulation package (VASP)~\cite{Kresse1993, Kresse1996a}. The kinetic energy cutoff of the plane wave basis for the DFT calculations was chosen to be 460 eV. A $\Gamma$-centered $4\times8\times 4$ $k$-point grids are used to perform the momentum-space calculations for the  Brillouin zone (BZ) integration. We have constrained the magnetic moments without incorporating SOC and switching off the crystal symmetry to capture the sole non-relativistic effect. A penalty contribution to the total energy is considered in these calculations to constrain the moments. The penalty energy fixes the local moment into a specific direction~\cite{Kresse1993, Kresse1996a}
\begin{equation}
E=E_0+\sum \gamma [\Vec{M}_i -\hat{M}^0_i(\hat{M}^0_i \cdot \Vec{M}_i) ]^2~,
\end{equation}
where $E_0$ is the DFT energy without any constraint, and the second term represents the penalty energy contribution due to non-colinear direction constraint. $\hat{M}^0_i$ and $\Vec{M}_i$ represent the unit vector along the desired direction of magnetic moment and integrated magnetic moment inside the Wigner-Seitz cell at site $i$, respectively. The sizes of the Wigner size radii are to be chosen as 1.98~\AA, 1.21~\AA, 1.09~\AA~and 0.24~\AA~for Ce, Ni, As and O atoms, respectively.  The choice of $\gamma$ controls the penalty energy contribution. In our calculations for CeNiAsO, we set $\gamma=6$ eV/$\mu_B^2$ to get negligible penalty contribution 1$\times$~10$^{-8}$ eV. We construct the tight-binding model Hamiltonian of CeNiAsO by using atom-centred Wannier functions within the VASP2WANNIER90 \cite{Souza2001} codes. Utilizing the obtained tight-binding model, we calculate the Edelstein response tensor.

\section{Acknowledgement}
JS, AC, ABH, and LS acknowledge funding by the Deutsche Forschungsgemeinschaft (DFG, German Research Foundation) - TRR 173 – 268565370 (project A03) and TRR 288 – 422213477 (project A09 and B05), and the Alexander von Humboldt Foundation.
TJ acknowledges support by the Ministry of Education of the Czech Republic CZ.02.01.01/00/22008/0004594 and, ERC Advanced Grant no. 101095925.
We acknowledge the high-performance computational facility of supercomputer `Mogon' at Johannes Gutenberg Universit\"{a}t Mainz, Germany. The authors acknowledge fruitful discussions with Gerrit Bauer, Rafael Gonz\'{a}lez Hern\'{a}ndez, and Nayra A Alvarez Pari.

\end{document}